\documentstyle[aps,graphics,multicol,psfig,epsfig,preprint,color]{revtex}  

\begin{document}

\tightenlines

\begin{center}

{\large {\bf Geometry of compact tubes and protein structures  }} \\

\vskip 2.0cm {\sc  Jayanth R.  Banavar$^{\dag,*}$, Alessandro Flammini$^{\ddag}$, 
Davide Marenduzzo$^{\ddag}$, Amos Maritan$^{\ddag,*}$ \& Antonio Trovato$^{\S}$
  }

\vskip 2.0cm

\end{center}

$^{\dag}$  Department of Physics,  104 Davey
Laboratory, The Pennsylvania  State  University,  University Park,  Pennsylvania
16802, USA \\

$^{\ddag}$  International  School for Advanced Studies  (S.I.S.S.A.), Via Beirut
2-4,  34014  Trieste,  INFM  and  the  Abdus  Salam  International   Center  for
Theoretical Physics, Trieste, Italy \\

$^{\S}$ Niels Bohr Institutet, Blegdamsvej 17, 2100 K{\o}benhavn {\O},
Denmark \\

* To whom correspondence should
be   addressed   
(E-mail:  banavar@psu.edu  and maritan@sissa.it).

\newpage

\parindent 0.0cm

{\bf   Proteins form a very important class of polymers.  In spite
of major advances in the understanding of polymer science, the
protein problem has remained largely unsolved.  Here, we show
that a polymer chain viewed as a tube not only captures
the well-known characteristics of polymers and their phases
but also provides a natural 
explanation for many of the key features of protein
behavior.  There are two natural length scales 
associated with a tube subject to compaction -- the
thickness of the tube and the range of the attractive
interactions.  For short tubes, when these length scales become
comparable, one obtains  marginally compact structures,
which are  relatively few in number  
compared to those in the generic compact phase of polymers.
The motifs associated with the structures in
this new phase include helices, hairpins and sheets. We suggest that Nature has
selected this phase for the structures of proteins 
because of its many advantages 
including the few candidate strucures, the ability to squeeze the
water out from the hydrophobic core and the flexibility and
versatility associated with being marginally compact.
Our results provide a framework for understanding the common 
features of all proteins. } \\

\newpage
A  revolution \cite{Watson} in the understanding of biomolecular
structure took place about 50 years ago using {\em the precise
geometrical
relationships among the atoms and molecules and the rigorous
application of the new structural principles}  enunciated by
Linus Pauling \cite{PaulingNobel}.  Earlier, in 1939, J. D. Bernal
\cite{Bernal}  had noted that 
{\em the symmetry of protein crystals is much higher than would be
expected statistically from compounds of such great complexity.
This would seem to indicate that each molecule is built of
subunits, themselves unsymmetrical but arranged in a symmetrical
way.}  The protein folding problem, the determination of the
structure of the folded state of a protein from knowledge of the
sequence of amino acids, has remained unsolved despite a large 
amount of experimental information on protein structures, the
availability of powerful computers and detailed knowledge of the
building blocks of proteins and their chemistry.  The approach
pioneered by Pauling is exceedingly effective for understanding 
small scale structures in great detail but becomes harder to
apply at the scale of full protein structure.  The complexity 
arises from the 20 types of naturally occurring amino acids
and the solvent and their mutual interactions. \\

An independent  approach to the study of such complex problems
consists of stepping back and adopting a coarse-grained view which
incorporates just the most essential elements which capture the important
emergent features.   For example, given a specific chemical
compound, one may use the  principles of quantum
mechanics and chemistry to determine its crystalline structure.
Alternatively, in a coarse-grained sense, one may deduce the
existence of several types of crystalline structure based on
general symmetry and packing considerations.  In this case, the specific 
chemistry of a material would dictate
which one of
these candidate structures the material would adopt.  
A common example is the
face-centered-cubic (fcc) arrangement adopted both by common salt
with strong electrostatic interactions and hard spheres (oranges
packed by a grocer)
in order to achieve the most efficient packing.  \\

In order to attack a problem of the magnitude of the structure of
complex biomolecules, we suggest that it is necessary to consider
both points of view.  Even assuming that computational power in
the future reaches a point which allows a brute force solution
incorporating all details, one might be able to mimic Nature but not necessarily
understand her.
On adapting a statement by Pauling \cite{PaulingC} (we have added three
words of our own), {\em the problem has been examined, in the
main, from one point of view only -- not the wrong point of view,
but one which, unaided, gives a vista insufficient to reveal the}
simplicity underlying the {\em true complex nature}.\\  

The power of the
coarse-grained approach is illustrated by considering some of the
familiar states of matter.  The gas, liquid and solid phases can
be understood in terms of atoms and their interactions
modelled as hard spheres (there is no distinction between the gas
and liquid phases in the absence of any attractive interaction) or
Lennard-Jones systems. On varying the nature of
interactions  and
thermodynamic quantities such as temperature and/or pressure, it
is possible to obtain   these states of
matter, which arise from the
collective, emergent behavior of a large number of  atoms. 
Glassy behavior ensues when the crystallization is thwarted by
dynamical constraints.  Entirely new classes of behavior
are found on considering 
anisotropic molecules as in liquid crystals \cite{Chaikin} -- the breaking of
the symmetry of the building blocks introduces qualitatively new features.\\ 

The molecules of life are chain molecules, polymers,  
which introduce the  feature
of connectivity along the chain.  Indeed, this feature has been
exploited by Nature in
the DNA molecule to code for genes.  
Detailed studies of polymers have revealed several phases
including a swollen phase (analogous to the gas phase)
corresponding to self-avoiding conformations, a highly
degenerate compact phase
in which different monomers of a chain have an effective
attraction to form a dense globule and semi-crystalline phases
\cite{Atlas}. \\ 

Proteins are an important class of chain biomolecules made up of 
amino acids.   These molecules fold into a somewhat compact state
with the folded structures controlling their functionality. 
The folding is driven by hydrophobic interactions or the tendency of 
certain amino acids to avoid water.
The structures of folded proteins  do not
correspond to the generic compact phase of a polymer.  First, the
total number of protein folds is only a few thousand
\cite{Chothia} instead of
the innumerable compact conformations of a generic chain of the
same length and second, the building blocks of biomolecular structure 
are pretty motifs including helices and sheets.  Indeed, generic
compact conformations are neither suitable for encoding specificity 
nor are they dynamically accessible in a simple manner. Furthermore,
proteins, while stable and able to fold rapidly and reproducibly to their
native state structures \cite{Anfinsen}, are  sensitive to the right
types of perturbations and are consequently 
able to perform a dizzying array of functions.  Here, we shall
argue that this phase adopted by molecules embodying life is a
new one distinct from the well-studied polymeric phases.\\

Let us begin with a set of unconstrained hard spheres with an attraction
of a given range.  When the attraction has a range smaller than
the diameter of the hard sphere, it is ineffective
and  the ground state is a  gas.  On increasing the range to a
value equal to   the hard sphere
diameter, the ground state changes from a gas to a
fcc crystal -- each hard sphere is surrounded by the maximum
possible number of other hard spheres.  Consider now a polymer chain made up of
hard spheres tethered together.  When the attraction between the
hard spheres is sufficiently short range,
the analog of the gas phase is a swollen phase corresponding to
self-avoiding conformations of the chain.  On turning on the
attraction by increasing its range, there is a change in the
ground state structure. 
One now gets a fcc lattice but this
time with the tethers running through the hard spheres.  There
are a huge number of ways of doing this and so the ground state
is highly degenerate.   Of course, a plethora of other compact states 
would be obtained if this
crystallized state is dynamically inaccessible due to the tethering or other
constraints.    Such compact states  are not good candidates 
for structures of proteins because there
are too many of them and accessing 
a specific structure is next to impossible.  Of course, 
constraints on the local curvature  of the chain could lead to 
other structures, such as lamellar semi-crystalline phases \cite{Atlas}, at
low temperatures.  \\

The model of spheres tethered together to form a chain does
capture the notion of connectivity of a chain but it leaves out a
second key factor, the inherent anisotropy associated with the
local directionality of the chain.  In other words, for a given 
sphere, the adjacent spheres along the chain define a local
direction adopted by the chain.  This anisotropy is most easily 
captured by replacing the sphere by an object with axial
symmetry, the simplest example of which is a coin or a circular
disc.  Indeed, an object made up of tethered coins of identical
size resembles a tube of uniform thickness such as a garden
hose. Such a tube like geometry is a
coarse-grained representation of the well-known steric effects
\cite{Rama} of
the basic constituents of proteins.   One may then ask what the analog 
of the fcc crystal phase is for such a tube. \\ 

There are two physical consequences of the tube picture.  First,
there is a local curvature constraint that does not permit radii
of curvature smaller than the tube thickness.  This feature may
equally well be captured for the case of spheres tethered
together by an explicit bond-bending energy term.  The second
consequence, which is unique to the tube description, is the 
inherent anisotropy which is reflected in the dependence of the 
potential energy of interaction between two spatially nearby
segments of the tube, not only on their distance from each other
but also on their mutual orientation  (Figure 1).  
This change in the
symmetry of isotropic interactions between spheres (that is
conventionally considered in the well-studied polymer case) to the 
cylindrical symmetry associated with the tube leads to qualitative
changes in the nature of the ground state conformations.
At high temperatures, for a tube, one expects a swollen
phase in which the local directions defining the tube are
distributed isotropically.  At sufficiently low temperatures, there
is a spontaneous symmetry breaking leading to the selection of a
preferred tube direction.  This transition is analogous to the
isotropic-nematic transition in liquid crystals \cite{Chaikin}
for tubes.  For  tube segments, which are thin compared to the range of the
attractive interaction and which are near each other, there is a
larger flexibility in their relative orientation (see Figure 1). 
Thus, in this case,  a dense globule phase with no significant orientational
ordering is expected at intermediate temperatures
between the swollen and the orientationally
ordered phases.\\

We now turn to the ground state structures of a simpler situation,
one exploited by Nature in
proteins, of a short tube when  the forces  promoting its compaction
just set in.  Indeed the interaction range and the 
tube size are matched for proteins because,  on the one hand, the
effective interactions in the presence of the solvent are
short  range  and the  squeezing  out of  water is
facilitated  by the outer atoms of nearby  side  chains  coming  together
and, on the other, it is these same side chains that determine the 
effective thickness of the tube.   Detailed  analytic and
numerical calculations show that one obtains far fewer conformations than
the corresponding generic polymer chain or a thin tube (allowing a protein
only relatively few selection choices for its native state
conformation) because the different parts of the tube have to 
position themselves just right relative to each other in order
to respect the inherent anisotropy and yet avail of the effective attraction
(Figure 1). \\

Consider the effect
of tuning the thickness of the short tube for a given range of
attraction.  When the tube thickness is bigger than the range of
attraction, one obtains a swollen phase because the attraction is
ineffective.   The other extreme is when the tube thickness is
very small compared to this range.  In this limit,  the
greater degrees of freedom for the relative positioning of nearby
tube segments due to the longer range of attraction leads to many
degenerate conformations.  \\

On varying the tube thickness,
near the point when the attractive forces have just set in, 
as described below,
different segments
of the tube have to position themselves just right with respect
to each other in order to avail of this attraction.  This has two
important consequences: first, the tube compaction
leads to the formation of a hydrophobic core in
the interior of the folded structure and, second, 
this careful relative positioning of the tube segments
combined with the anisotropy associated with the tube 
weeds out all but a few  from the list of possible
candidate structures for ground state conformations.  These
conformations may be thought of as being marginally
compact and are  thus
attractive candidates for versatility and flexibility, because
they are able to respond to small changes in an effective
manner.    Furthermore, on lowering the temperature, starting
from the swollen phase, one would expect an almost immediate
ordering at a relatively low temperature (with respect to a
thinner tube) into one of the ground state conformations without any
partially folded intermediates.  This ``two-state" character
is an important feature of small globular proteins
and arises in the tube context  because the scale
of the interaction energy goes down as the tube becomes thicker,
entropic effects are less important at the low temperatures of
ordering  and
the orientational effects become stronger (see Figure 1). \\ 

We show below that the building blocks of these structures are the
familiar helix, hairpins and sheets.   Furthermore, 
elementary considerations predict the geometry of an ideal helix,
which is very close to that observed in Nature, and the zig-zag
appearance of the strands. \\ 

We have carried out computer simulations and analytic
calculations of short tubes in the
marginally compact phase.  The resulting structures are shown in
Fig. 2.
Helices and hairpins (sheets) are of course the well-known building blocks of
protein structures \cite{Pauling1,Pauling2} (see Fig. 2 (A1) and (D1) for 
two examples from a protein and (A2), (D2) and (D3) for the tube
structures in our simulations). In addition to the prediction of
these motifs in our calculations, it is interesting to note that
some of the other marginally compact conformations bear a qualitative resemblance
to secondary folds in biopolymers.   Helices analogous to Fig. 2 (A3)
with an irregular contact map occur, e.g., in the HMG protein NHP6a \cite{hmg}
with pdb code 1CG7.  Fig. 2 (C1) shows the ``kissing hairpins'' \cite{kiss}
of RNA (pdb code 1KIS), each of which is a distorted and twisted hairpin 
structure while Fig. 2 (C2) is the corresponding tube
conformation.  
Fig. 2 (B1)  shows a helix of strands
found experimentally in Zinc metalloprotease \cite{zinc} (pdb code: 1KAP), whereas
Fig. 2 (B2) is the corresponding marginally compact conformation obtained 
in our calculations. \\

It is possible to understand the results shown in Fig. 2 by
means of simple arguments.   Let us begin by 
taking a piece of  tube of radius $R_0$ and length equal to $2 \pi
R_0$.  Gonzalez and Maddocks  \cite{K1} have shown that a simple
description of a tube is obtained by taking all triplets
of points along the axis of the tube and measuring the radii of
the circles passing through them 
with a view of ensuring that none of these radii is smaller than
the thickness.  In particular, the local radius of curvature of a
tube can never be smaller than its thickness.
By placing the tube in the form of a donut of radius $R_0$, one can
effectively fill all the space in the middle of the donut.
When the tube is longer than $2 \pi R_0$, the most efficient means
of compactifying it is to place it in a helical conformation
with local radius of curvature
equal to $R_0$ and with the pitch chosen so that
the segments of the tube in successive turns lie on top of each other.
This is, of course, a valid structure only when the range of
attractive interactions allows contacts to be made in this
geometry.
This ideal space-filling helix has  a special pitch to radius
ratio (see Fig. 2 (A2)),  which is observed not only in
$\alpha$-helices in globular proteins but also in the helices of collagen
\cite{Maritan}.   An effective squeezing out of the space
between the successive turns of the helix is accomplished by the fact that
the orientations of the interacting segments of the tube
are parallel to each other.  Were this not
to be the case, the inherent anisotropy of the tube (imagine a tube made up of
discrete coins) would lead to a mismatch,
a factor of no consequence in a chain made up of tethered hard
spheres.  
One may show analytically that, on increasing the tube thickness,
helices
are excluded from being the ground states,
when the tube thickness exceeds
$R_0^{max,hel}\sim\left (\sqrt{1+R_1^2}\right)/2\sim 0.943$
($R_1$ is the range of the attractive interaction and is chosen to be 1.6
units in Fig. 2 -- all lengths in the simulation are measured in units of
the distance between successive $C_{\alpha}$ atoms) which is
obtained when two parallel straight lines (successive turns of the helix
treated as circles with infinite
radius) are 
at a distance of
$R_1$ from each other.  Indeed this structure is one that
corresponds to a hairpin. \\

The zig-zag hairpin of Fig. 2 (D2) is a distorted version of this
idealized case due to the discreteness of the protein chain. 
One can use elementary geometrical considerations to
prove that the zig-zag nature accommodates a tube of larger
thickness compared to straight segments.
For two zig-zag antiparallel strands facing each other, one can show
analytically that the maximum thickness is obtained (leaving aside
the edge effect of how the
strands are connected together in a hairpin) 
when one has a space-filling conformation.
Indeed, this condition leads to the following relationship between the tube
thickness $R_0$ and the interaction range $R_1$
$$
R_1^2+2+\frac{R_1}{R_0}-4R_0^2=0     ,
$$
which yields a value of $R_0 \sim 1.2124$, when $R_1 =1.6$, in
perfect accord with our simulations. \\

For intermediate tube thicknesses between those corresponding to
a helix or a hairpin,  we find only a few other structures that 
may be thought of as interpolating between the two
limiting cases.   In order of decreasing thickness, one obtains
first the kissing hairpin structure (Fig. 2 (C2)), which is a hairpin
twisted into three dimensions -- a feature allowed for by the
slightly smaller thickness compared to the planar hairpin; a
helix made up of strands (Fig. 2 (B2)); and irregular, somewhat non-ideal
helices (Fig. 1 (A3)).  In all cases, nearby parts of the tube
are oriented parallel to each other. \\ 

It is interesting to consider the ground state of many long tubes
subject to compaction.  Packing considerations suggest that the
tubes become essentially straight and parallel to each other 
and are arranged (when viewed
end on) in a triangular lattice, analogous to the Abrikosov flux
lattice phase in superconductors \cite{Tinkham}.   Returning to
the case of a single tube, in the very long length limit, a
similar phase would be expected with the additional constraint of
the bending of the tube segments at the ends.  As stated
before, for a discrete chain, a planar placement of zig-zag
strands is able to accommodate the largest thickness tube that can yet
avail of the attraction -- however, the thickness for this
limiting case is too large to produce the three dimensional
ordering alluded to above.  It would be interesting to consider
how the ground state structure crosses over from the
``flux-lattice" type phase to the familiar planar phase.  Indeed,
for thick tubes of moderate length, one may expect to form a
large sheet-like structure analogous to the
cross-$\beta$-scaffold observed as a building block of amyloid
fibrils \cite{Dobson}.  Such fibrils have been implicated in a 
variety of human disorders
including Alzheimer's disease and spongiform encephalopathies 
such as Creutzfeldt-Jakob disease.   Remarkably, recent findings
suggest that the ability of proteins to form amyloid
is {\em a generic property of polypeptide chains} \cite{Dobson}. \\

Many strategies for attacking the protein folding problem have
been put forward which employ a coarse-grained description
\cite{Editorial}.  None
of the currently used methods has been successful.  Our results
suggest that a deficiency of all these methods has been that the context
provided by the local tube orientation is neglected while
considering the interaction between coarse-grained units.
The novel phase discussed here arises from the addition of
anisotropy to the well-studied polymer problem just as one
obtains rich liquid crystal behavior on studying anisotropic
molecules.  A mapping of the phase behavior of tubes
on varying the nature of interactions, the thickness of
the tube, the length of the tube 
and temperature might yield additional surprises. \\

In 1939, J. D. Bernal \cite{Bernal} wrote: {\em Any effective picture of
protein structure must provide at the same time for the common
character of all proteins as exemplified by their many chemical
and physical similarities, and for the highly specific nature of
each protein type.}  Our results provide a simple framework for the
common character of all proteins.  Our analysis is based on
just three ingredients -- all proteins share a backbone, 
there are effective forces which promote the folding of a protein
and the one and only new idea that a protein can be viewed as a tube
(see Fig. 3).   We have not introduced any input into our
analysis which pertains to {\em the highly specific nature of each
protein type} \cite{Bernal} as encoded by the amino acid
sequence.   It would be interesting to extend our calculations to
a tube of non-uniform thickness.  For example, the presence of a
small amino acid like glycine at backward bends allows for tight
turns to be formed to facilitate good packing
and lead to low values of local thickness. 
Also, the wide variety of amino acid properties such as
hydrophobicity, charge and ability to form disulfide or hydrogen
bonds may be captured in a coarse-grained way by inhomogeneous
attractive amino acid specific interactions, which respect the 
inherent anisotropy of a tube. \\

It is
important to stress that our results are not at odds with or meant as
a substitute for the detailed and beautiful work 
involving the laws of quantum mechanics and biochemistry.  The virtue 
of our approach is that it predicts
a novel phase with selected types of structures and the attendant
advantages.   It is then necessary to complement this information with
the principles of quantum chemistry to assess whether a given
biomolecule would fit one of these structures.  We do
not invoke hydrogen bonds as Pauling did in his prediction of 
protein secondary motifs \cite{Pauling1,Pauling2}
and indeed not all the structures in the marginally compact phase are
compatible with hydrogen bond placement.   
What is remarkable, however,  is that the lengths 
of the covalent and hydrogen  bonds and the rules of quantum chemistry
conspire to provide a perfect fit to the basic structures in this novel phase.  
One cannot but be amazed at how the evolutionary forces of Nature have
shaped the molecules of life \cite{Denton} ranging from 
the DNA molecule, which carries the genetic
code and is efficiently copied, to
proteins, the work horses of life, whose functionality follows from 
their form which, in turn, is a novel phase of matter. \\

{\bf Acknowledgements} 
We are indebted to  Flavio Seno and Michele
Vendruscolo for useful discussions.
This  work  was supported by INFM, MURST cofin2001, NASA and
the Penn State MRSEC under NSF grant 
DMR-0080019.  
\\

\newpage

FIGURE CAPTIONS

Figure 1:

Potential energy of interaction of two straight tubes 
as a function of their mutual distance and relative orientation.

The top panel shows the simplified geometry that we have considered.
Two straight  tubes, each of length $ 2 l$,  are placed a distance $d$ from each
other  with their axes making an angle $\theta$ with respect to
each other.  The line joining the centers of the tubes is perpendicular
to both the tube axes.  We consider a favorable energy of interaction when a
pair of infinitesimal segments of the axes of the two tubes 
are within  a distance $R_1$ (chosen
to be 1.6 units as in the simulations described in Fig. 2), 
which is the range of a uniform attractive interaction.  The lower panel shows
plots of the potential energy both as a function of $d$ and $\theta$ for
$l = 1$.   The left hand figure shows how the tube geometry leads to an
anisotropic interaction, reflected by an energy which depends on $\theta$, for three
values of $d$.  Note that in each case the
energy has been scaled by the energy when the tubes are parallel to each
other with the corresponding value of $d$.  
The anisotropy becomes more pronounced as the tubes become thicker because
this restricts the possible range of $d$ to values closer to $R_1$.
The weak minimum for the
$d = 0$ case away from $\theta =0$ is due to the short length of the tubes.
The lower right hand panel shows  a plot of the magnitude of the potential
energy when the tubes are oriented parallel to each other as a function of $d$.
(we have chosen units such that the scale of the attractive
interaction energy of two segments within the range of attraction is simply
given by the product of their lengths.)
The potential energy is zero when the value of $d$ exceeds that of $R_1$.
Note that, for a continuum tube,
as the tube thickness, $R_0$,  increases towards $R_1/2$,
restricting $d$ to values close to but smaller than $R_1$,
there are two simultaneous effects.  First the scale of the interaction energy
becomes very weak and second, the anisotropy becomes pronounced.  
As described in the text, both these effects play a crucial role in simplifying
the behavior of proteins. \\

Figure 2:

Building blocks of biomolecules and ground state
structures associated with
the marginally compact phase of a short tube. \\

In order to mimic a protein, the axis of the tube of non-zero
thickness (radius of cross-section) $R_0$ is modelled
as a one dimensional discrete chain, whose bonds are
of fixed length (set equal to 1 without loss of generality -- all 
other lengths will be measured in these units from now on) and
which connect neighboring $C_{\alpha}$ atoms along the chain.  
The thickness \cite{K1} of the 
tube is captured by disallowing conformations for which
$R_0>{\rm min}_{i \neq j \neq k}R_{i,j,k}$, where $R_{i,j,k}$ is the 
radius of the circle going through the centers of the atoms $i$, $j$ and 
$k$:
$$
R_{i,j,k}= \frac{r_{i,j} r_{j,k} r_{i,k}}{4 A_{i,j,k} }
$$
where $A_{i,j,k}$ is the area of the triangle through $i$, $j$ and $k$ 
and $r_{i,j}$ is the 
distance between the centers of the $i$-th and the $j$-th atoms.
Indeed, one may ascribe a local thickness to the tube by
measuring all three body radii associated with a given atom and
all other pairs and selecting the smallest radius among these.
Figure 3 shows the distributions of the local tube thickness for
the native state structures of 30 proteins and underscores the excellent
approximation of viewing the protein as a tube of uniform
thickness. \\

The interaction between non-consecutive atoms 
is modeled via a 2-body potential with a hard core and a square well:
$$
V(r_{i,j})= 
\left \{ 
\begin{array}{cc}  
 \infty & \mbox{ if $r_{i,j} < 2  R_{h.c.}$ } \\
 -1   & \mbox{ if $ 2 R_{h.c.}< r_{i,j} <  R_{1}$ } \\
  0   & \mbox{ if $R_{1} < r_{i,j}  $ }
\end{array}
\right.
$$
The three-body interactions capture the inherent anisotropy of a
tube, whereas the pairwise potential drives the compaction.
For the results shown here, 
$ R_{h.c.}$ has been set to $0.55$, $R_1$ to $1.6$ and
$R_0$ was increased in the vicinity of the transition to the swollen
phase until the number of pairwise contacts was reduced to 
three.  
While these values have been selected in order to
mimic the protein backbone formed by the $C_{\alpha}$ atoms,
we have verified that our results are
robust to variations in these values. \\

The top row shows some of the building blocks of biomolecules, while the
second row depicts the corresponding structures obtained for a
tube. (A1) is an $\alpha$-helix of a naturally occuring protein, while
(A2) and (A3) are the helices obtained in our calculations -- (A2)  has
a regular contact map and is obtained when $R_0 = 0.80267 $ whereas (A3) 
($R_0 =   0.833 $) is a distorted helix
in which the distance between successive atoms along the helical axis
is not constant but has period $2$. (B1) is a helix of strands in the alkaline protease
of pseudomonas aeruginosa, whereas (B2) shows the corresponding structure
($R_0 = 0.88 $)
obtained in our computer simulations. 
(C1) shows the ``kissing'' hairpins of RNA
and (C2) the corresponding conformation obtained in our simulations
with $R_0=0.95$. Finally (D1) and (D2)
are two instances of quasi-planar hairpins.
The first structure is from
the same protein as before (the alkaline protease of pseudomonas aeruginosa) while
the second is a  typical conformation found in our simulations when
$R_0>0.98$.  The sheet-like structure (D3) is obtained for a
longer tube. \\

Figure 3:

Distribution of local thicknesses of the native state structures
of 30 proteins.  The peaked distribution shows that it is a 
good approximation to
think of a protein as a tube of uniform thickness of
around $2.7 \AA$.

\newpage

\begin{figure}   
{\includegraphics[width=6.in]{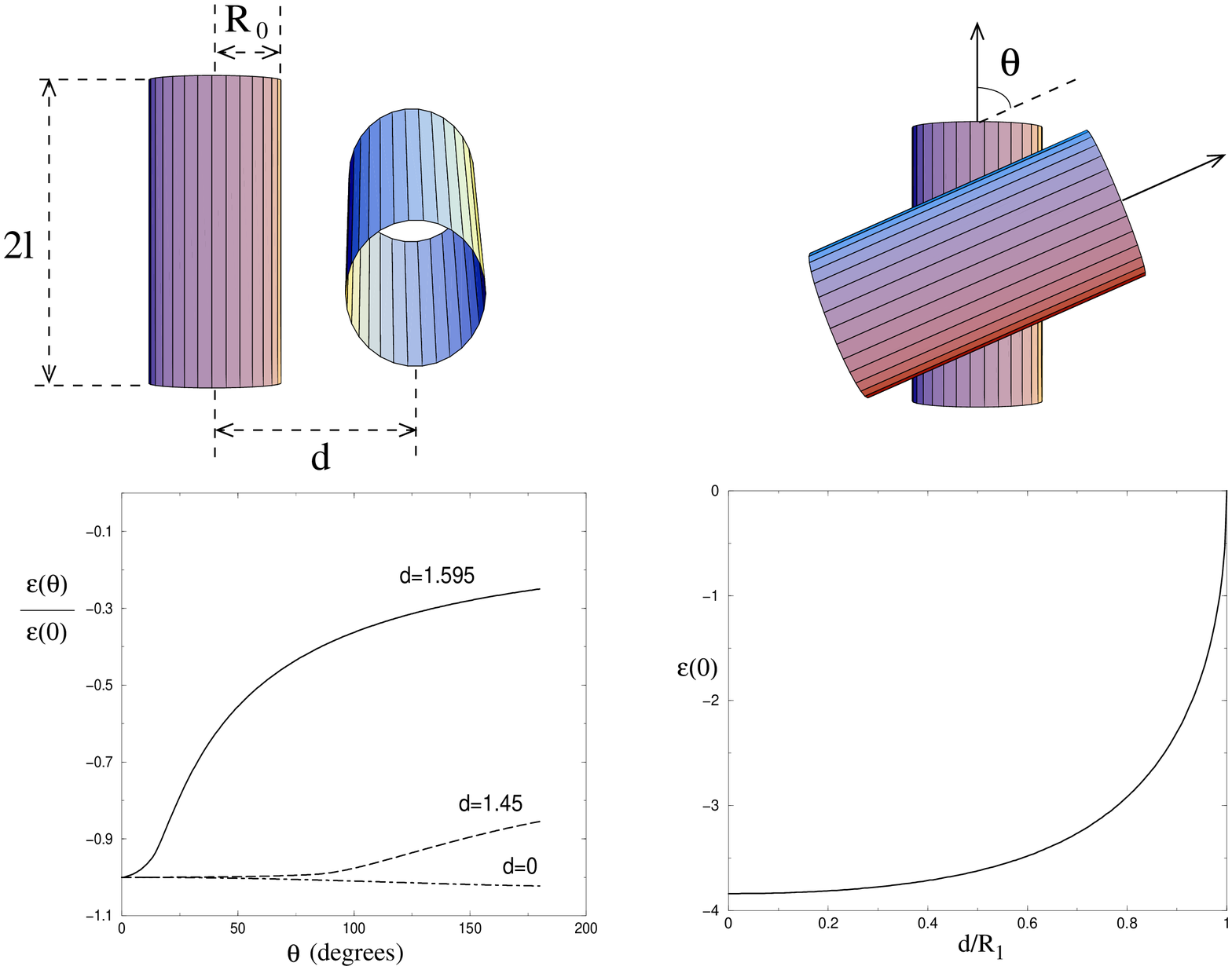}}  
\end{figure}

%\begin{figure}
%{\includegraphics[angle=270,width=6.in]{figure1.eps}}
%\end{figure}

\newpage

\begin{figure}
{\includegraphics[angle=270,width=6.in,clip]{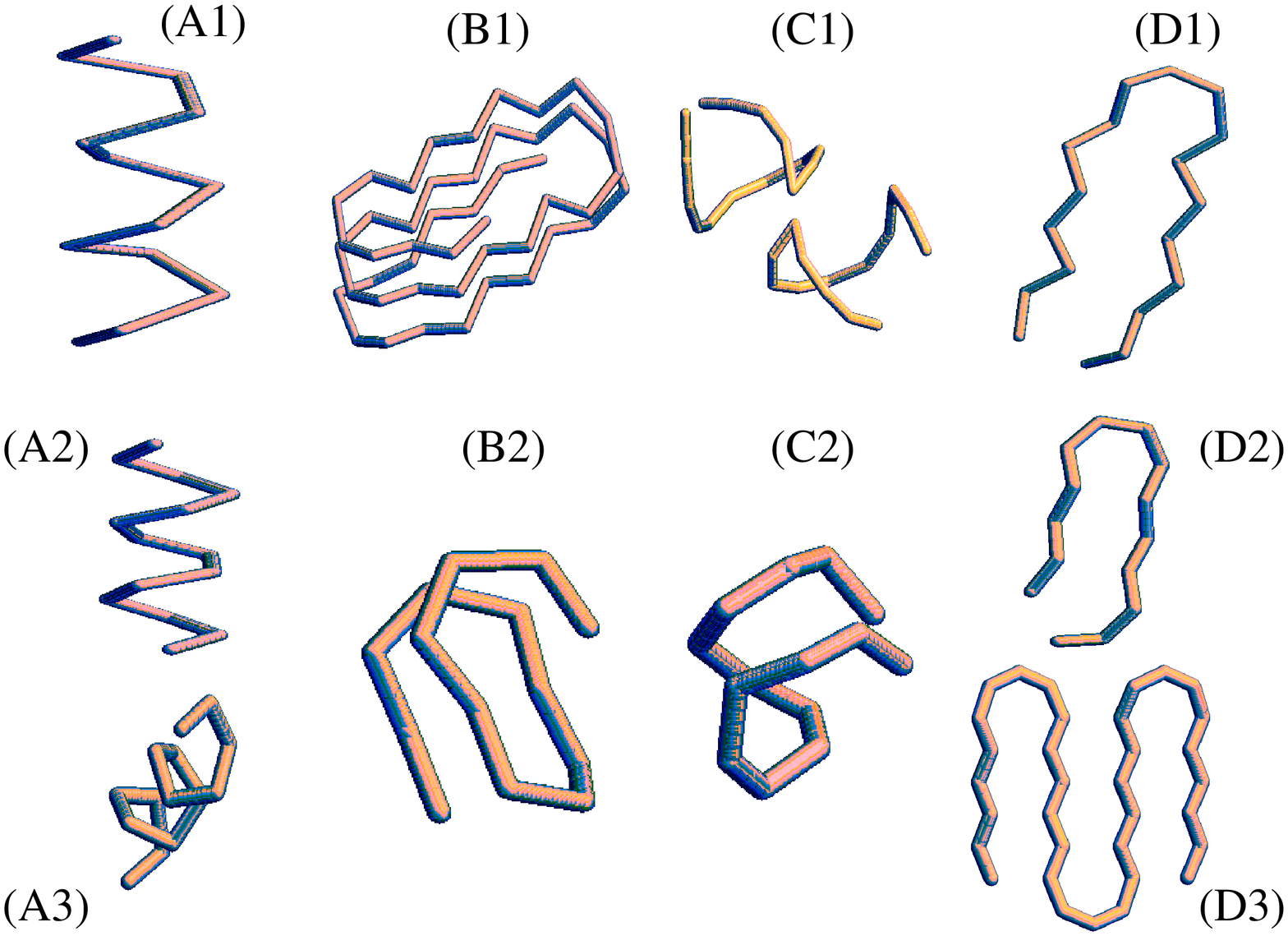}}
\end{figure}

\newpage    
\begin{figure}
{\includegraphics[angle=270,width=6.in,clip]{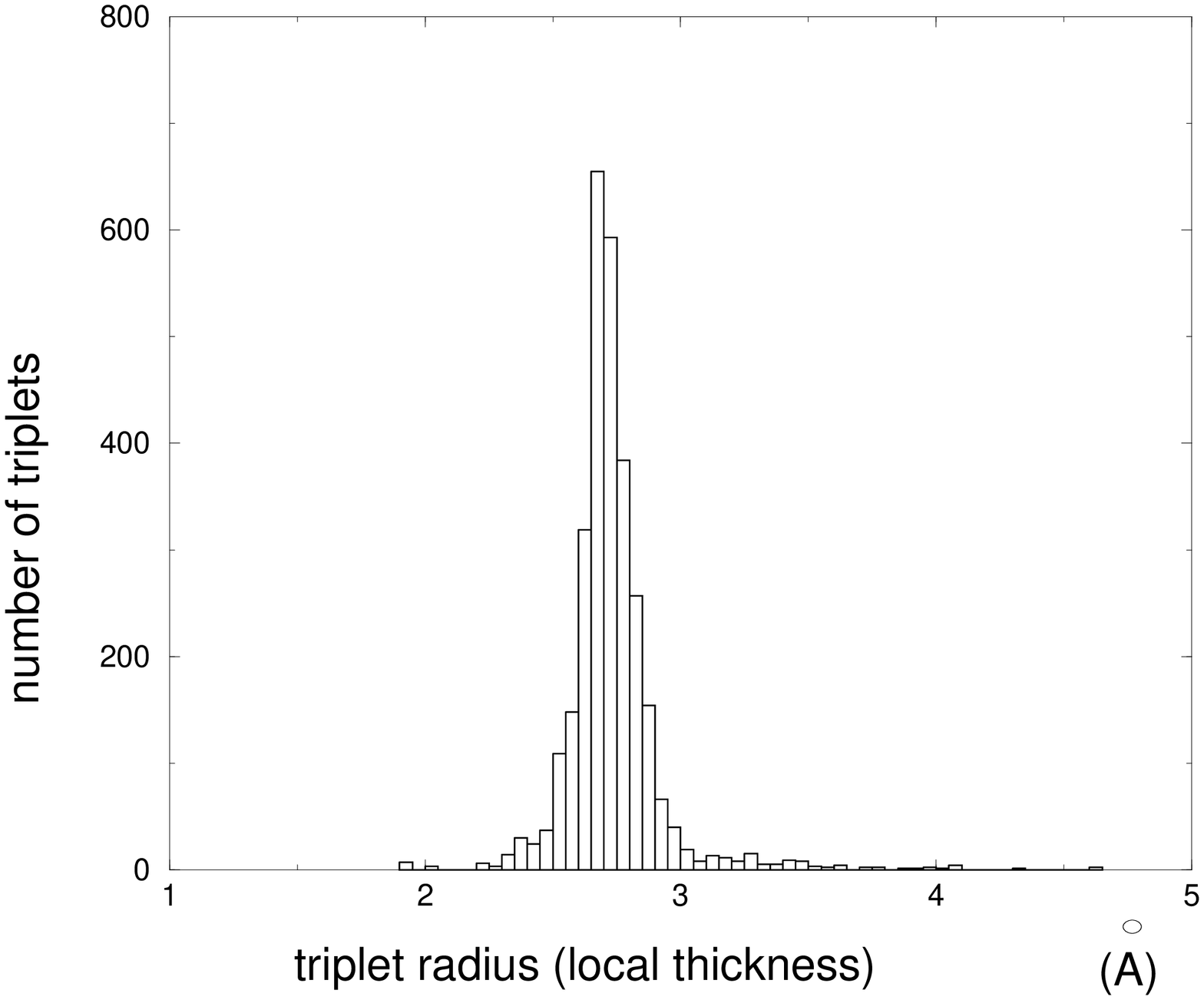}}
\end{figure}    

\end{document}